\documentclass[reprint,superscriptaddress,nofootinbib,amsmath,amssymb,physrev]{revtex4-2}
\pdfoutput=1
\usepackage{float}
\usepackage{xcolor}
\usepackage{graphicx}
\usepackage{dcolumn}
\usepackage{bm,lipsum}
\usepackage{siunitx}
\usepackage[colorlinks=true, linkcolor=blue, citecolor=blue, urlcolor=blue]{hyperref}

\newcommand{\ZIB}{Zuse Institute Berlin, 14195 Berlin, Germany}
\newcommand{\TUB}{Technische Universität Berlin, 10623 Berlin, Germany}
\newcommand{\FUB}{Freie Universität Berlin, 14195 Berlin, Germany}

\usepackage{amsmath}
\usepackage{graphicx}
\usepackage{booktabs}
\usepackage{tabularx}
\usepackage{xcolor,colortbl}
\usepackage[T1]{fontenc}

\newcommand{\mytoprule}{\specialrule{0.1em}{0.2em}{0.2em}}
\newcommand{\mymidrule}{\specialrule{0.1em}{0.2em}{0.2em}}
\newcommand{\mybottomrule}{\specialrule{0.1em}{0.2em}{0.2em}}

\definecolor{Gray}{gray}{0.9}

\newcolumntype{a}{>{\columncolor{Gray}}c}
\newcolumntype{b}{>{\columncolor{white}}c}

\begin{document}

\title{AAA rational approximation for scatterometry of nanoscale gratings}
\author{Livius~Leibl}
\affiliation{\FUB}
\author{Viktoriia~Sytenko}
\affiliation{\TUB}
\author{Rana~Demircioglu}
\affiliation{\TUB}
\author{Leonid~Beliaev}
\affiliation{\FUB}
\author{Felix~Binkowski}
\affiliation{\ZIB}

\begin{abstract}
We investigate the applicability of AAA rational approximation to scatterometry-related response functions of nanoscale gratings. A silicon line grating with material dispersion is considered. We compare the convergence of AAA rational approximation with that of cubic spline interpolation as the number of frequency samples is increased. The results demonstrate the potential of rational approximation, consistent with the pole structure of the considered response function arising from resonance effects of the grating.
\end{abstract}

\maketitle
\section{Introduction}
Scatterometry is a widely used optical technique for the characterization of nanoscale gratings~\cite{Madsen_2016}. It is based on measuring the optical response of a sample and relating the measured data to mathematical models of the investigated sample~\cite{Gross_2006}. This enables the characterization of parameters such as feature dimensions, periodicity, and material properties. One approach is spectroscopic ellipsometry~\cite{Fujiwara_2007}, which provides frequency-dependent information on changes in the polarization state of reflected light and has been applied for the characterization of gratings~\cite{Raymond_1995,Kumar_Scatterometry_2014,Wurm_2017,Mukherjee_2025}.

In numerical simulation frameworks for scatterometry applications~\cite{Kato_2012}, the optical response is typically obtained from computationally demanding electromagnetic simulations using, e.g., finite-difference time-domain (FDTD) methods~\cite{Yee_FDTD_1966}, rigorous coupled-wave analysis (RCWA)~\cite{Hugonin_Reticolo_2025}, or finite-element methods (FEM)~\cite{Monk_2003}. Evaluating the response functions over a frequency range can therefore require substantial computational effort. Thus, an accurate approximation of the response functions from a limited number of frequency samples can reduce the required computational effort.

Resonance effects are exploited to improve the sensitivity of scatterometry setups~\cite{Coulombe_1999,Antos_2005,Mukherjee_2025}, where the corresponding optical response functions exhibit poles in the complex frequency plane~\cite{Binkowski_PRB_2024}. These poles can be attributed to the resonances of the nanostructures, which are solutions to the source-free Maxwell equation with open boundary conditions~\cite{Lalanne_QNMReview_2018,Lalanne_QNM_Benchmark_2018}. Rational approximation of the response functions is particularly well suited in this case, as rational functions can naturally represent the poles. The AAA algorithm~\cite{Nakatsukasa_2018}, pronounced triple-A and named after adaptive Antoulas--Anderson~\cite{Antoulas_1986}, is a powerful approach for rational approximation and has recently attracted significant attention in numerical mathematics and various application fields~\cite{Nakatsukasa_2025}. It has also been applied to investigate resonance phenomena in nanophotonic systems~\cite{Betz_LPOR_2024,Betz_AAA_BranchPoints_2025,Bruno_AAA_2026}.

In this work, we investigate the applicability of AAA rational approximation to scatterometry-related response functions of nanoscale gratings. A silicon (Si) line grating with dispersive material properties is considered. We apply the AAA algorithm to two different contexts: (i)~constructing a generalized Lorentz model to describe the material dispersion of Si, using the AAA algorithm to obtain an initial rational approximation of the material dispersion; and (ii)~computing a rational approximation of the response function under investigation. For the latter case, the convergence of AAA rational approximation is compared with that of cubic spline interpolation as the number of frequency samples is increased. We demonstrate superior convergence behavior of AAA rational approximation, which can be attributed to the pole structure of the considered response function. We relate these poles to resonance effects of the grating by solving the source-free Maxwell equation, yielding the underlying eigenvalues and eigenmodes.

\begin{figure}
\includegraphics[width=0.49\textwidth]{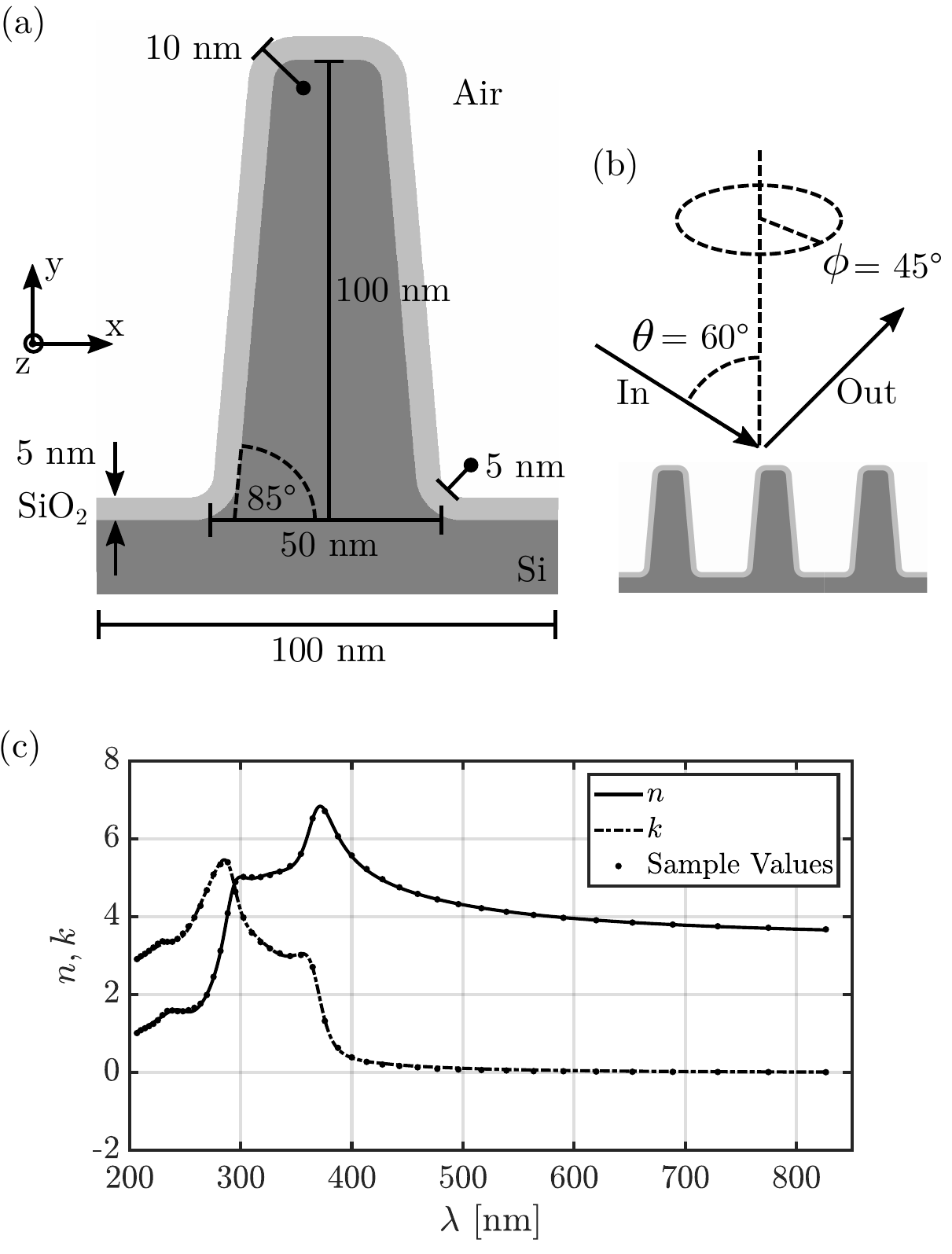}
\caption{\label{figure1}
Nanoscale grating based on Ref.~\cite{Wurm_2017}. 
(a)~Periodic unit cell of the Si line grating under investigation. The grating is considered as an open system, and it is surrounded by a silicon dioxide (SiO$_2$) layer with a refractive index of $n=1.5$. 
(b)~Scatterometry setup with incoming and outgoing plane waves. The direction of the illumination is defined by the angles of incidence $\theta$ and $\phi$. 
(c)~Refractive index $n(\lambda)$ and extinction coefficient $k(\lambda)$ of the considered Si grating as a function of the wavelength $\lambda$ of the plane wave illumination. The lines represent the generalized Lorentz model given by Eq.~\eqref{eq:lorentz}. The dots are the $46$ sample values taken from the tabulated data in Ref.~\cite{Aspnes_1983}.
}
\end{figure}

\section{AAA rational approximation}
The effectiveness and broad applicability of the AAA algorithm rely on the combination of a numerically stable barycentric representation and a greedy support-point selection strategy~\cite{Nakatsukasa_2018,Nakatsukasa_2025}. The AAA algorithm computes a rational approximation of a function $f(z)$. The inputs are sample points $z_{j} \in Z \subseteq \mathbb{C}$ in the complex plane, together with the corresponding sample values $f_j = f(z_j)$, and the output is a rational approximation in barycentric representation,
\begin{equation} \label{eq:AAA_rat_approx}
   \left. f(z) \approx \frac{n(z)}{d(z)} = \sum^n_{j=0} \frac{\beta_j f_j}{z-t_j} \middle/ \sum^n_{j=0} \frac{\beta_j}{z-t_j}. \right.
\end{equation}
The set $Z$ is chosen in the region of physical interest.
The barycentric support points $t_j$ are greedily added to $\hat{Z} \subset Z$ until a desired target accuracy is reached. The barycentric weights $\beta_j$ minimize the error in a least-squares sense,
\begin{equation}
    \sum_{z_j \in Z \setminus \hat{Z}} |f_j d(z_j) - n(z_j) |^2, \nonumber
\end{equation}
where $\sum_{j=0}^n |\beta_j|^2 = 1$.
The poles $\tilde{z}_k$ of the rational approximation are the zeros of the denominator $d(z)$, and they are given by the eigenvalues of the $(n+2)\times(n+2)$ generalized eigenproblem
\begin{equation} \label{eq:eig_rat_approx}
\begin{bmatrix}
0 & \beta_{0} & \beta_{1} & \cdots & \beta_{n} \\
1 & t_{0} & & & \\
1 & & t_{1} & & \\
\vdots & & & \ddots & \\
1 & & & & t_{n}
\end{bmatrix}
x = \lambda
\begin{bmatrix}
0 & & & & \\
& 1 & & & \\
& & 1 & & \\
& & & \ddots & \\
& & & & 1
\end{bmatrix}
x. \nonumber
\end{equation}
For a simple pole $\tilde{z}_k$, using the Laurent expansion around $\tilde{z}_k$, the residues $a_k$ of the rational approximation can be computed by the formula
\begin{equation}
    a_k = \frac{n(\tilde{z}_k)}{d^{\prime}(\tilde{z}_k)}, \nonumber
\end{equation}
where the denominator $d(z)$ is differentiated with respect to $z$. This gives the residue-based expansion
\begin{equation}
    f(z) \approx a_0 + \sum^{n}_{k=1} \frac{a_k}{z-\tilde{z}_k}, \nonumber
\end{equation}
where $a_0$ is a normalization constant. 

Note that residues and poles are computed at the end of the approximation process, as reflected by the fact that they do not appear explicitly in the barycentric representation given by Eq.~\eqref{eq:AAA_rat_approx}.

\section{Scatterometry of a nanoscale grating}
For the characterization of nanoscale structures, the steady-state scattering response of an open system can be described by the time-harmonic Maxwell equation equipped with open boundary conditions,
\begin{align}
	\nabla \times  \mu_0^{-1} 
	\nabla \times \mathbf{E}(\mathbf{r},\omega)  -
	\omega^2\epsilon(\mathbf{r},\omega) \mathbf{E}(\mathbf{r},\omega)  = 
	i\omega\mathbf{J}(\mathbf{r}), \label{eq:maxwell}
\end{align}
where $\mathbf{E}(\mathbf{r},\omega)\in\mathbb{C}^3$ denotes the electric field, $\mathbf{J}(\mathbf{r})\in\mathbb{C}^3$ is the impressed current source, $\omega\in\mathbb{C}$ is the angular frequency, and $\mathbf{r}\in\mathbb{R}^3$ represents the spatial coordinate. The material dispersion and the spatial distribution of material are characterized by the permittivity $\epsilon(\mathbf{r},\omega)=\epsilon_0\epsilon_\mathrm{r}(\mathbf{r},\omega)$, where $\epsilon_\mathrm{r}(\mathbf{r},\omega)$ is the relative permittivity and $\epsilon_0$ is the vacuum permittivity. At optical frequencies, magnetic effects are typically negligible, such that the permeability $\mu(\mathbf{r},\omega) = \mu_\mathrm{r}(\mathbf{r},\omega) \mu_0$ equals the vacuum permeability $\mu_0$.

In the following, we investigate a nanoscale grating introduced in Ref.~\cite{Wurm_2017}, shown schematically in Fig.~\ref{figure1}(a). The parameters of the grating are based on those reported in Ref.~\cite{Wurm_2017} and are modified for illustrative purposes. The grating is illuminated by plane waves with fixed angles of incidence and varying frequency $\omega$. The setup is sketched in Fig.~\ref{figure1}(b). We consider the ratio
\begin{align}
    \rho(\omega)=\frac{J_{\mathrm{pp}}(\omega)}{J_{\mathrm{ss}}(\omega)}, \label{eq:rho}
\end{align}
where $J_{\mathrm{pp}}(\omega)$ and $J_{\mathrm{ss}}(\omega)$ denote the diagonal entries of the Jones matrix for $p$- and $s$-polarized illumination, respectively~\cite{BornWolf_PrinciplesOptics}.
The complex-valued quantity $\rho(\omega)$ is directly related to scatterometry setups, which characterize the amplitude ratio and phase difference between the $p$- and $s$-polarized components of the reflected light.
For a given $\omega$, Eq.~\eqref{eq:maxwell} is solved twice, once for each incident polarization. The corresponding reflected far-field amplitudes obtained from the solutions $\mathbf{E}(\mathbf{r},\omega)$ are used to construct the Jones matrix. The solution of Eq.~\eqref{eq:maxwell} and the subsequent post-processing are performed using the FEM solver JCMsuite~\cite{Pomplum_NanoopticFEM_2007}.

Note that the Jones matrix is generally not diagonal, and polarization conversion between $p$- and $s$-polarized components may occur. For this numerical study, we consider the ratio of the diagonal Jones-matrix elements as a representative target quantity.

\setlength{\tabcolsep}{10pt}
\renewcommand{\arraystretch}{1.5}
\renewcommand{\arraystretch}{1.5}
\begin{table}[t]
\caption{Parameters of the four-pole generalized Lorentz model for Si given by Eq.~\eqref{eq:lorentz}. The residues $a_k$ and poles $\tilde{\omega}_k$ are given in units of $10^{15}\,\mathrm{s}^{-1}$, and $\epsilon_\infty$ is dimensionless.}
\vspace{0.25cm}
\centering
\begin{tabular}{c c c}
\mytoprule
\multicolumn{3}{c}{$\epsilon_\infty = 0.1584$}\\
\mymidrule
$k$ & $a_k$ & $\tilde{\omega}_k$ \\
\mymidrule
1 & $3.588 + 3.572i$      & $5.116 - 0.1961i$ \\
2 & $0.2103 + 24.34i$     & $6.012 - 0.8841i$ \\
3 & $1.273 + 6.180i$      & $6.447 - 0.2655i$ \\
4 & $3.235 + 0.005778i$   & $7.573 - 0.6022i$ \\
\mybottomrule
\end{tabular}
\label{tab:lorentz_parameters}
\end{table}

\subsection{Material model}
The permittivity of Si is approximated by a rational function, which enables the construction of a physically consistent model for the material dispersion. We apply the AAA algorithm to compute such a rational approximation, using sample points and sample values taken from Ref.~\cite{Aspnes_1983}. To ensure causality, we then remove the contributions corresponding to the poles located in the upper half of the complex frequency plane. We further enforce Hermitian symmetry, $\epsilon_\mathrm{r}^*(\omega)=\epsilon_\mathrm{r}(-\omega)$, which ensures that the corresponding time-domain response function is real-valued. Targeting a maximum relative error in the magnitude of the model of less than $5\,\%$ yields a four-pole generalized Lorentz model for the material response,
\begin{align} 
\epsilon_\mathrm{r}(\omega) = \sum_{k=1}^4 \left[\frac{i a_k}{\omega-\tilde{\omega}_k} + \frac{i a_k^*}{\omega+\tilde{\omega}_k^*}\right]+\epsilon_{\infty}, \label{eq:lorentz}
\end{align}
where $a_k$ and $\tilde{\omega}_k$ denote the modified residues and poles, respectively. The values of the material model parameters are given in Table~\ref{tab:lorentz_parameters}. Although passivity is not explicitly enforced, the model satisfies the passivity condition. A comprehensive study of the construction of material models using the AAA algorithm is in preparation~\cite{Barkova_AAA_Material_Model}.

Figure~\ref{figure1}(c) compares the refractive index $n(\omega)$ and the extinction coefficient $k(\omega)$ obtained from the proposed model with the tabulated material data from Ref.~\cite{Aspnes_1983}, where $\epsilon_\mathrm{r}(\omega)  = (n(\omega)+ik(\omega))^2$. Note that, in the following, results are visualized as functions of the wavelength $\lambda = 2 \pi c/\omega$, following common practice.

\subsection{Rational approximation}
\begin{figure}
\includegraphics[width=0.49\textwidth]{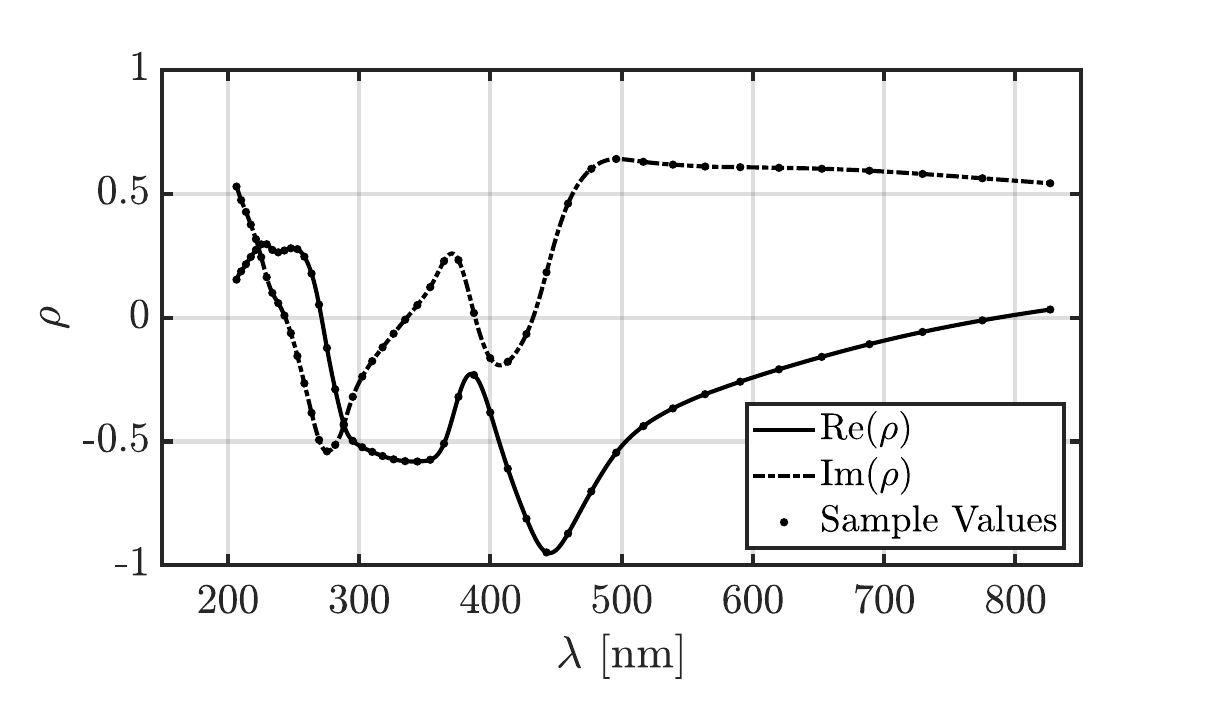}
\caption{\label{figure2}
AAA rational approximation of $\rho(\lambda)$ based on all $46$ available sample points $\lambda_k$ from the tabulated data in Ref.~\cite{Aspnes_1983}. The corresponding sample values $\rho_k$ are indicated by dots.
}
\end{figure}

The AAA algorithm is applied to construct a rational approximation of $\rho(\lambda)$ given by Eq.~\eqref{eq:rho}. We solve Eq.~\eqref{eq:maxwell} using the $46$ sample points $\lambda_k$ and corresponding Si data from Ref.~\cite{Aspnes_1983}, yielding the sample values $\rho_k=\rho(\lambda_k)$. The AAA algorithm then uses $\lambda_k$ and $\rho_k$ as input and yields an approximation with a maximum relative error in the magnitude of $6.8\times10^{-15}$. The approximation is constructed from $24$ support points and $23$ poles, corresponding to the highest possible degree for the given number of sample points~\cite{Nakatsukasa_2018}. The rational approximation of $\rho(\lambda)$ and the sample values are shown in Fig.~\ref{figure2}.

Although the approximation is constructed from only $24$ support points, the AAA algorithm uses all $46$ sample values during the approximation process. The selected support points determine the rational representation, while the remaining sample points are used to minimize the approximation error during the adaptive selection process. Therefore, the accuracy at the original sample points does not provide information about the approximation quality at unsampled wavelengths.

\begin{figure*}
\includegraphics[width=0.98\textwidth]{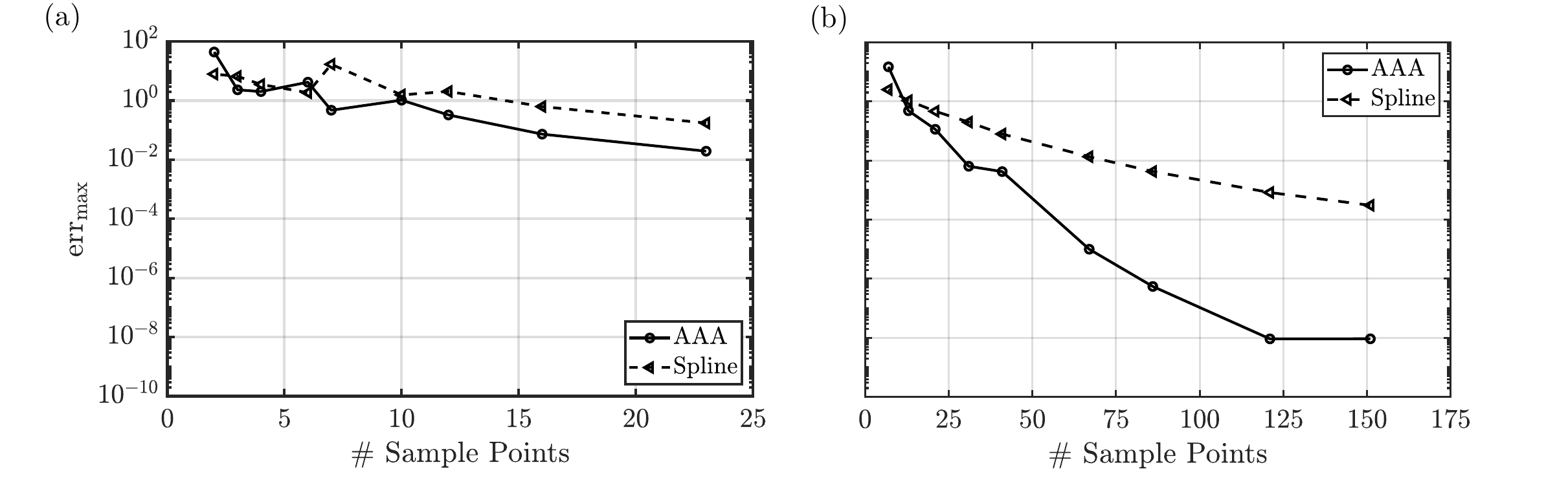}
\caption{\label{figure3}
Convergence of $\rho(\lambda)$ with respect to the number of used sample points $\lambda_k \in \Lambda$ for AAA rational approximation and for cubic spline interpolation. The maximum relative error $\mathrm{err}_\mathrm{max}$ is computed by Eq.~\eqref{eq:relerror}.
(a)~Sample points correspond to the tabulated data from Ref.~\cite{Aspnes_1983}, where the reference set $\Lambda_\mathrm{ref}$ of wavelengths comprises $46$ points. This tabulated data is also used for describing the material dispersion of Si.
(b)~Sample points are equispaced in the wavelength interval $200\,\mathrm{nm}<\lambda_k<800\,\mathrm{nm}$, where $\Lambda_\mathrm{ref}$ comprises $601$ points with a spacing of $1\,\mathrm{nm}$. The generalized Lorentz model given by Eq.~\eqref{eq:lorentz} is used for describing the material dispersion of Si.
}
\end{figure*}

\subsection{Convergence study}
To investigate the approximation error at unsampled points, we consider a reference set $\Lambda_\mathrm{ref}$ of wavelengths.
A subset $\Lambda \subset \Lambda_\mathrm{ref}$ is selected as input, together with the corresponding sample values $\rho_k$ for $\lambda_k \in \Lambda$. The resulting approximation is evaluated over $\Lambda_\mathrm{ref}\setminus\Lambda$ and compared with the reference values. This procedure is performed for different subsets $\Lambda$, where the subsets are obtained by successively decreasing the sampling step size. The approximation quality is assessed at the points $\Lambda_\mathrm{ref}\setminus\Lambda$, which are not provided to the algorithm. The maximum relative error is defined as
\begin{equation}
\mathrm{err}_\mathrm{max} =
\max_{\lambda \in \Lambda_\mathrm{ref}\setminus\Lambda}
\left(
\frac{\left|\rho_\mathrm{approx}(\lambda)-\rho_\mathrm{ref}(\lambda)\right|}
{\left|\rho_\mathrm{ref}(\lambda)\right|}
\right), \label{eq:relerror}
\end{equation}
where $\rho_\mathrm{approx}(\lambda)$ denotes the approximation obtained from the sample points $\lambda_k \in \Lambda$, and $\rho_\mathrm{ref}(\lambda)$ denotes the reference values.

We compare AAA rational approximation with cubic spline interpolation, 
a commonly used approach for approximating wavelength-dependent quantities in computational photonics. Figure~\ref{figure3}(a) shows the convergence of the approximations of $\rho(\lambda)$ using only the wavelengths available from the tabulated data in Ref.~\cite{Aspnes_1983}. Due to the limited number of available sample points and potential measurement noise in the tabulated data, only a moderate reduction of the approximation error can be achieved, resulting in a maximum relative error of about $2\,\%$ for the AAA algorithm and $17\,\%$ for cubic spline interpolation.

Figure~\ref{figure3}(b) shows the convergence of the approximation of $\rho(\lambda)$, where the generalized Lorentz model from Eq.~\eqref{eq:lorentz} is used for the material dispersion of Si. This allows for arbitrarily fine sampling of $\rho(\lambda)$, where equispaced sampling is used~\cite{Huybrechs_AAA_2023}. For $13$ or more sample points, the AAA algorithm yields a more accurate approximation than cubic spline interpolation. Using $121$ sample points, the AAA algorithm yields an approximation with a maximum relative error of $9.2\times10^{-9}$, while cubic spline interpolation yields an error of $8.3\times10^{-4}$. The overall approximation accuracy is limited by the numerical accuracy of the Maxwell solver.

The superior convergence of AAA rational approximation observed can be attributed to the fact that the optical response function under investigation exhibits poles in the complex plane associated with resonance effects. Rational functions provide a natural representation of such functions. In contrast, polynomial-based interpolation methods do not explicitly capture pole-like behavior.

\subsection{Low-dimensional model and poles}
\begin{figure}
\includegraphics[width=0.49\textwidth]{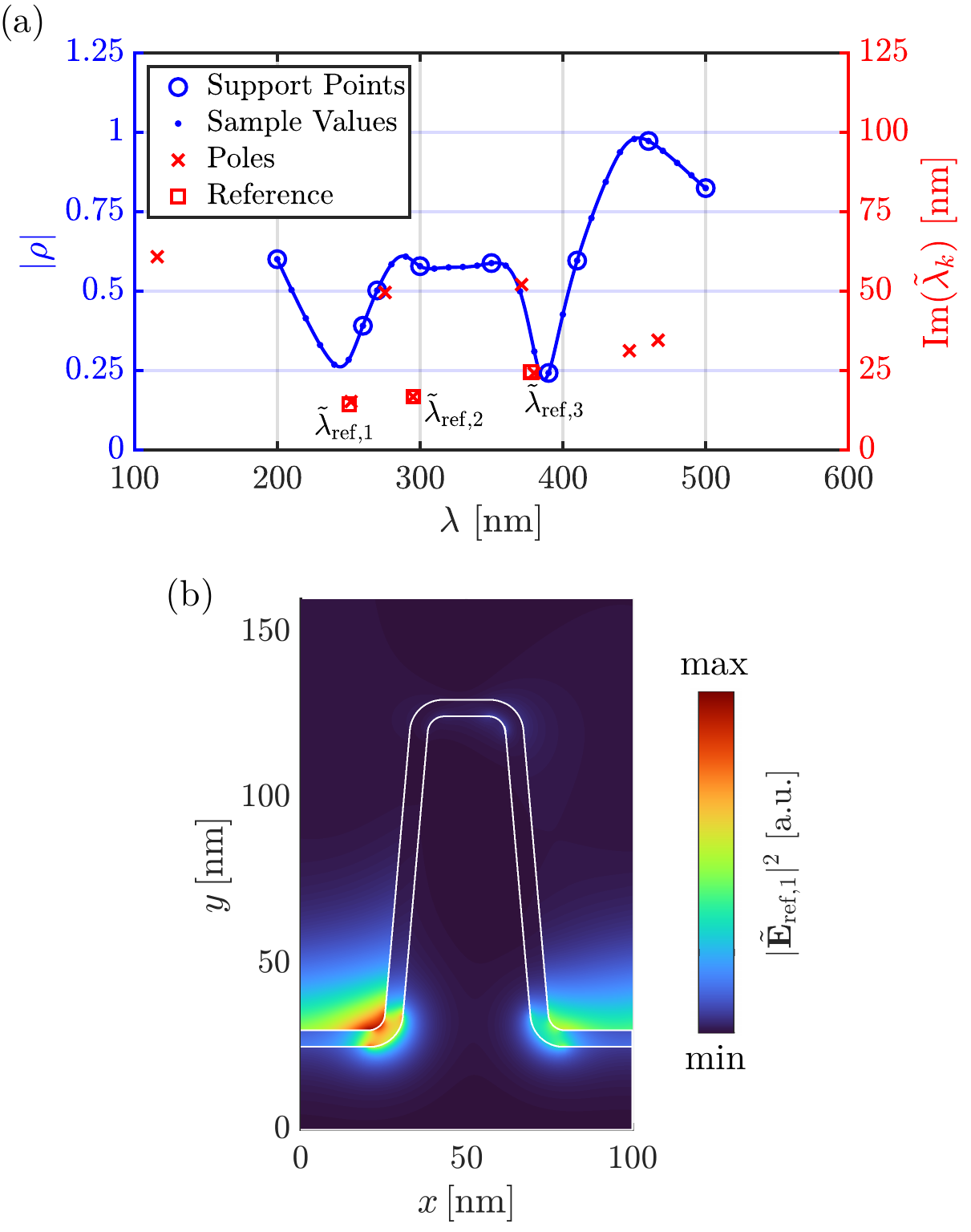}
\caption{\label{figure4}
Low-dimensional model of $\rho(\lambda)$ with corresponding poles in the complex plane.
(a)~AAA rational approximation based on eight poles $\tilde{\lambda}_k$ yields a maximum relative error in $|\rho(\lambda)|$ of about $0.5\,\%$. The reference solutions $\tilde{\lambda}_\mathrm{ref,k}$ are computed by solving eigenproblems using the source-free form of Eq.~\eqref{eq:maxwell}.
(b)~Electric field intensity $|\tilde{\mathbf{E}}_\mathrm{ref,1}|^2$ of the eigenmode $\tilde{\mathbf{E}}_\mathrm{ref,1}$ corresponding to $\tilde{\lambda}_\mathrm{ref,1}$.
}
\end{figure}

Rational models are characterized by their poles. For the convergence study shown in Fig.~\ref{figure3}, the tolerance of the AAA algorithm is set to $10^{-13}$ to ensure sufficiently accurate approximations. Alternatively, one can set a higher tolerance to obtain a model with a smaller number of poles. Figure~\ref{figure4}(a) shows such an approximation, in which $\rho(\lambda)$ is represented by a rational function with eight poles $\tilde{\lambda}_k$ and a maximum relative error of about $0.5\,\%$, where $31$ equispaced sample points are used.

To associate these poles with resonances of the grating, we compute reference solutions for the poles by solving eigenproblems, i.e., the source-free form of Eq.~\eqref{eq:maxwell}, using the Arnoldi algorithm~\cite{Saad_Book_NumMeth_Eig_2011}. This yields the reference eigenvalues $\tilde{\lambda}_\mathrm{ref,k}$ and eigenmodes $\tilde{\mathbf{E}}_\mathrm{ref,k}$.
A fixed-point iteration is applied, using the poles obtained from the AAA algorithm as initial guesses. We compute reference solutions only for poles that remain nearly unchanged when varying the tolerance of the AAA algorithm. This criterion is used to distinguish between stable poles associated with resonance effects and numerical poles that cannot be interpreted physically but improve the accuracy of the approximation. These reference solutions $\tilde{\lambda}_\mathrm{ref,k}$ are also shown in Fig.~\ref{figure4}(a). It can be observed that the reference solutions lie close to the poles of the AAA rational approximation.

Figure~\ref{figure4}(b) shows the electric field intensity of the eigenmode $\tilde{\mathbf{E}}_\mathrm{ref,1}$, which is localized at the rounded corners of the grating at the substrate interface. This eigenmode corresponds to a resonance of the grating. The eigenmodes $\tilde{\mathbf{E}}_\mathrm{ref,2}$ and $\tilde{\mathbf{E}}_\mathrm{ref,3}$ are not shown because their electric field intensities are localized predominantly in the Si substrate, i.e., they can be interpreted as numerical modes based on the numerical realization of the open boundary conditions.

Note that the rational material model provides an analytical continuation into the complex plane, which also allows for the use of complex-valued sample points. This could further improve convergence of the AAA rational approximation~\cite{Betz_LPOR_2024}.

We further note that the wavelengths considered in this study are larger than the periodicity of the grating. Therefore, branch points in the complex plane do not have to be considered. In Ref.~\cite{Betz_AAA_BranchPoints_2025}, AAA rational approximation is used to investigate the interplay between such scattering thresholds and resonances.

\section{Conclusion}
A Si line grating with material dispersion was studied numerically based on the scatterometry-related ratio $\rho(\lambda)$ given by Eq.~\eqref{eq:rho}. The applicability of AAA rational approximation to this optical response function was investigated. It was demonstrated that rational approximation converges faster than cubic spline interpolation with respect to the number of wavelength samples. This behavior can be attributed to resonance effects of the grating, which are naturally represented by the poles of rational functions.

To further investigate this connection, a low-dimensional rational model with eight poles was constructed, yielding an approximation with a maximum relative error of about $0.5\,\%$. It was shown that a certain pole identified by the AAA algorithm can be associated with a resonance of the system.

\section*{Data availability} 
Source code and simulation results for the numerical experiments for this work can be found in the open access data publication~\cite{Leibl_SourceCode_AAA_Grating}.

\section*{Conflicts of interest} 
The authors declare no conflicts of interest.

\section*{Acknowledgments}
We gratefully acknowledge discussions with Matthias Wurm, Anna Barkova, Fridtjof Betz, Martin Hammerschmidt, Lin Zschiedrich, and Sven Burger. 

This work was carried out within the Student Research Opportunities Program (StuROPx) of the Berlin University Alliance. We acknowledge funding by the Deutsche Forschungsgemeinschaft (DFG, German Research Foundation) under the Excellence Strategy of the Federal and State Governments via the Berlin University Alliance and via the Berlin Mathematics Research Center MATH+ (EXC-2046/2, project ID: 390685689). We further acknowledge funding by the German Federal Ministry of Research, Technology and Space (BMFTR, Forschungscampus MODAL, project 05M20ZBM).

\end{document}